\documentclass[12pt,letterpaper]{article}
\usepackage{amsmath,amssymb}
\usepackage{graphicx,color}
\usepackage[breaklinks]{hyperref}

\usepackage[bf]{caption}
\renewcommand{\captionfont}{\small\itshape}

\newcommand{\rr}{\mathbb{R}}

\newcommand{\be}{\begin{equation}}
\newcommand{\ee}{\end{equation}}
\newcommand{\ba}{\begin{aligned}}
\newcommand{\ea}{\end{aligned}}
\newcommand{\ben}{\begin{displaymath}}
\newcommand{\een}{\end{displaymath}}
\newcommand{\bea}{\begin{eqnarray}}
\newcommand{\eea}{\end{eqnarray}}
\newcommand{\bean}{\begin{eqnarray*}}
\newcommand{\eean}{\end{eqnarray*}}

\newcommand{\p}{\partial}

\usepackage{amsmath,amssymb}
\usepackage{graphicx,color}

\def\r {\rho}

\def\L{\Lambda}
\def\th {\theta}
\def\a {\alpha}

\def\b {\beta}

\def\d {\delta}

\def\s {\sigma}

\def\m{\mu}
\def\n{\nu}
\def\k{\kappa}

\definecolor{green}{rgb}{0,0.5,0}

\def\p{\partial}

\long\def\symbolfootnote[#1]#2{\begingroup
\def\thefootnote{\fnsymbol{footnote}}\footnote[#1]{#2}\endgroup}

\begin{document}

\begin{titlepage}
\vspace{10pt} \hfill {HU-EP-12/19} \vspace{20mm}
\begin{center}

{\Large \bf Quantization of AdS$\times$S particle in static gauge}

%\vspace{1cm}

\vspace{45pt}

{George Jorjadze,$^{a,\,b}~$
Chrysostomos Kalousios,$^a$ Zurab Kepuladze,$^c~$
%\symbolfootnote[2]{\tt{\{jorj,ckalousi,\}@physik.hu-berlin.de}}
}
\\[15mm]

{\it\ ${}^a$Institut f\"ur Physik der
Humboldt-Universit\"at zu Berlin,}\\
{\it Newtonstra{\ss}e 15, D-12489 Berlin, Germany}\\[3mm]
{\it${}^b$RMI $\&$  Free University of Tbilisi,}\\
{\it Bedia Str., 0183, Tbilisi, Georgia}\\[3mm]
{\it${}^c$ITP, Ilia State  University $\&$ Andronikashvili  Institute of Physics, \\
Tamarashvili Str. 6, 0177, Tbilisi,  Georgia}

\vspace{20pt}

\end{center}

\vspace{40pt}

\centerline{{\bf{Abstract}}}
\vspace*{5mm}
\noindent
We quantize the particle dynamics in AdS$_{N+1}\times $S$^M$ spacetime in static gauge,
which leads to the coordinate representation with wave functions
depending only on spatial coordinates. The energy square operator is quadratic in canonical momenta
and contains a scalar curvature term. We analyze the
self-adjointness of this operator and calculate its spectrum. We then construct unitary representations
of the isometry group  SO$(2,N)\times$SO$(M+1)$ and calculate the quantum relation between the Casimir numbers.
\vspace{15pt}
\end{titlepage}

\newpage

\subsection*{Introduction}

In this paper we quantize the AdS$\times$S particle dynamics in static gauge. We use
the quantization scheme of \cite{Dorn:2010wt} based on a covariant construction
of the energy square operator in the coordinate representation, where the wave functions
depend only on spatial coordinates.
Let us first outline the scheme, which we present here in a slightly modified form.
%\vspace{3mm}

Particle dynamics in a spacetime with coordinates $x^\mu,$ $\mu=(0,1,...,{\mathcal N})$ and a metric tensor
$g_{\mu\nu}(x)$ can be described by the following action in the first order formalism
\be\label{action g}
S=\int {d}\tau \Big(p_\m\,\dot x^\m-\frac{\lambda}{2}\left(g^{\m\n}p_\m p_\n+{\mathcal M}^2\right)\Big)~.
\ee
Here ${\mathcal M}$ is the particle mass, $\lambda$ is a Lagrange multiplier and its variation provides
the mass-shell condition
\be\label{massshell}
g^{\m\n}p_\m p_\n+{\mathcal M}^2=0~.
\ee

Using the Faddeev-Jackiw reduction \cite{Faddeev:1988qp} in the gauge
\be\label{gauge}
x^0+p_0\tau=0~,
\ee
from \eqref{action g} we get an ordinary Hamiltonian
system\footnote{We neglect the total derivative term
$-\frac{\mathrm{d}}{\mathrm{d}\tau}\left(\frac{1}{2}\,p_0^2\tau\right)$.}
\be\label{action h}
S=\int {d}\tau \Big(p_n\,\dot x^n-\frac{1}{2}\,p_0^2\Big)~,
\ee
where $p_0^2$, as a function of the spatial coordinates and momenta $(x^n, p_n)$
($n=1,...,{\mathcal N})$,
is obtained from the constraint \eqref{massshell} and
the gauge fixing condition \eqref{gauge}.

Notice that here we have modified the form of the standard static gauge $x^0=\tau$ used
in \cite{Dorn:2010wt}. However, this modification does not change the quantization scheme
and it appears more convenient for a Hamiltonian treatment of \eqref{action h} in a static spacetime,
as well as for the generalization to string theory \cite{Jorjadze:2012iy}.

A static  spacetime metric tensor can be represented in the form
\be\label{metric g}
g_{\mu\nu}=\left(\begin{array}{cc}
              g_{00} & 0 \\
              0 & g_{mn}
            \end{array}\right)~,
\ee
where $g_{00}$ and $g_{mn}$ are functions only of the spatial coordinates $x^n$.
In this case the particle energy $E(p,x)=-p_0 >0$ is conserved and from \eqref{massshell} follows that
\be\label{E^2 g}
E^2=\L(x)\,g^{mn}(x)~p_mp_n+{\mathcal M}^2\,\L(x)~,
\ee
with $\L(x):=-g_{00}(x)>0$.

Thus, the Hamiltonian in \eqref{action h} corresponds to a motion of a particle in the potential field
$V(x)=\frac{1}{2}\,{\mathcal M}^2\,\L(x)$ and in a curved background with metric tensor
\be\label{x metric}
h_{mn}(x)=\frac{1}{\L(x)}\,\,g_{mn}(x)~.
\ee

It is  natural to quantize this system in the coordinate representation, where the wave
functions $\psi(x)$
form a Hilbert space with covariant scalar product
\be\label{scalar product}
\langle \psi_2 | \psi_1 \rangle =  \int d^{\mathcal N}x \,\sqrt{h(x)} ~ \psi_2^*(x) \, \psi_1(x)~,
\qquad  \qquad h(x):=\mbox{det}\,h_{mn}(x)~.
\ee

On the basis of DeWitt's construction for quadratic in momenta operators \cite{DeWitt-R},
it was argued in \cite{Dorn:2010wt} that the energy square operator is given by
\be\label{E^2 g op}
E^2=-\Delta_h+\frac{{\mathcal N}-1}{4{\mathcal N}}\,\mathcal{R}_h(x)+{\mathcal M}^2\,\L(x)~.
\ee
Here  $\Delta_h$ is the covariant Laplace-Beltrami operator for the metric tensor $h_{mn}$ and
$\mathcal{R}_h(x)$ denotes the corresponding scalar curvature. The solution of the eigenvalue problem
for \eqref{E^2 g op} then provides the energy operator in diagonal form.

The coefficient in front of the scalar curvature term has been a subject of discussions during decades
(see \cite{Bastianelli:2006rx} and references therein). Therefore it is useful to comment on the value
of this coefficient, $\frac{{\mathcal N}-1}{4{\mathcal N}}$, chosen in \eqref{E^2 g op}.

For the particle dynamics in AdS$_{{\mathcal N}+1}$ this coefficient
was calculated in \cite{Dorn:2010wt} from the commutation relations of the symmetry generators.
In this case $\mathcal{R}_h(x)$
corresponds to the curvature of a semi-sphere and, therefore, it is constant.
The obtained constant shift in the energy square operator provides its spectrum in the form $(E_0+n)^2$,
with fixed $E_0$ and a non-negative integer $n$, that leads to the correct energy spectrum for the AdS particle.

For a generic ${\mathcal N}+1$ dimensional static spacetime the same value of the coefficient
$\frac{{\mathcal N}-1}{4{\mathcal N}}$ follows from the equivalence
between the static gauge quantization
and the covariant quantization based on the Klein-Gordon type equation.

The covariant quantization is a more conventional approach to
the particle dynamics in AdS backgrounds \cite{Breitenlohner:1982bm, ads-review}.
The general case with arbitrary dimensions and radii in this approach was
analyzed in \cite{Dorn:2003au} from the perspective of scalar field propagators
(see also \cite{Dai:2009zg}).

An interesting alternative quantization scheme based on the twistor and BRST
formalisms was proposed in \cite{Claus:1999jj} for a massive bosonic particle in AdS$_5$.
The obtained twistor construction was related to the oscillator construction of
\cite{Gunaydin:1998sw}.

The quantization of a superparticle in the AdS$_5\times$S$^5$ background
was done in \cite{Horigane:2009qb}, using  the lightcone gauge and the technique developed in \cite{Metsaev:1999gz}
(see also \cite{Siegel:2010gm}, which is based on a different lightcone type gauge).

The dynamics of a massive bosonic particle in AdS$_5\times$S$^5$ was considered in \cite{Passerini:2010xc}.
The authors used gauge invariant approach\footnote{Other references on gauge
invariant quantization of the AdS particle dynamics one can find in \cite{Dorn:2010wt}.} and Dirac brackets formalism.
The obtained results were applied for the analysis of string energy spectrum
at large coupling in the context of the AdS/CFT correspondence.

The main motivation of most of papers on particles dynamics in the AdS spaces is to develop  useful methods
for the quantization of strings in these backgrounds.

Our motivation is the same and in this paper we aim to apply the static gauge quantization
to AdS$_{N+1}\times$S$^M$ particle.
$\mbox{AdS}_{N+1}$ will be realized as
a hyperbola $X^A\,X_A=-R^2,\,$ with $\,A=(0',0,1,...,N),$ embedded in $\rr^{2,\,N}$  and
$S^M$ as a $M$-dimensional sphere $Y_I\,Y_I=R_S^2$ in $\rr^{M+1},$ with $\,I=(1,...,M+1)$.

In the next section we deal with the classical case and prepare the system for quantization.

\subsection*{Geometry on AdS$_{N+1}\times$S$^M$ }

AdS$_{N+1}\times$S$^M$ has ${\mathcal N}=N+M$ spatial coordinates and we associate the first $N$
coordinates with the AdS part and the rest $M$ coordinates with S$^M$. The time coordinate $x^0$
is given by the polar angle $\th$ in the $(X^{0'},X^0)$ plane.

We parameterize the embedding coordinates of $\rr^{2,N}$  as follows
\be\label{coordinates}
X^{0'}=\frac{R}{\r}\,\,\sin\th~,\qquad X^{0}=\frac{R}{\r}\,\,\cos\th~,\qquad
X^a = \frac{R}{\r}\,\,x^a~,\qquad \r:=\sqrt{1-x^b x^b}~,
\ee
where the coordinates $x^a$ are given on the $N$-dimensional unit disk.
%\footnote{For notational convenience we write these coordinates with down indices.}

The choice of coordinates $\phi^\a$ on S$^M$ is not important for our calculations,
since the quantization of the spherical part is trivial.
We use the Greek letters $(\a, \b)$ for tensorial indices on S$^M$ and they run from $N+1$ to $N+M$.

The induced metric tensor on AdS$_{N+1}\times$S$^M$ has the structure \eqref{metric g} with
\be\label{induced metric}
g_{00}=-\frac{R^2}{\r^2}~,\qquad
g_{mn}=\left(\begin{array}{cc}
              g_{ab} & 0 \\
              0 & g_{\a\b}
            \end{array}\right)~,
\ee
where $g_{ab}$ corresponds to the spatial part of $\mbox{AdS}_{N+1}$
\be\label{metric on AdS}
 g_{ab}=\frac{R^2}{\r^2}\left(\d_{ab}+\frac{x^a\,x^b}{\r^2}\right)
\ee
and $g_{\a\b}$ is the metric tensor on S$^M$. Note that the scalar curvature for $g_{mn}$ is given by
\be\label{R_g}
{\mathcal R}_g=-N(N-1)\,R^{-2}+M(M-1)\,R_S^{-2}~.
\ee

It is worth mentioning that the metric tensor \eqref{metric on AdS} corresponds to the Euclidean AdS space
(Lobachevski plane) and after the rescaling \eqref{x metric} with the Weyl factor
\be
\frac{1}{\L(x)}=\frac{\r^2}{R^2}~,
\ee
it becomes the metric tensor on the unit semi-sphere (see Fig. 1) with
\be\label{u metric}
h_{ab}(x)=\d_{ab}+\frac{x^a\,x^b}{\r^2}~, \qquad \mbox{and} \qquad  \mbox{det}\,h_{ab}=\frac{1}{\r^2}~ .
\ee

The total background metric tensor $h_{mn}$ defined by \eqref{x metric} has a similar to
\eqref{induced metric} block structure and from \eqref{u metric} we find the integration measure
in \eqref{scalar product} to be
\be\label{det h}
\sqrt{h}=\k^{-M}\,\mu_S\,\,\r^{M-1}~,
\ee
where $\k=R/R_S$ and $\mu_S$ is the SO$(M+1)$ invariant measure on the unit sphere.

The Laplace-Beltrami operator for the metric tensor $h_{mn}$ then reads
\be\label{Laplace op}
\Delta_h=h^{ab}\,\p^2_{ab}-(N+M) x^a\,\p_a +\frac{\k^2}{\r^2}\,\,\Delta_S~,
\ee
where
\be\label{h^ab}
h^{ab}=\d_{ab}-x^a\,x^b
\ee
is the inverse to  $h_{ab}$
and $\Delta_S$ is the Laplace-Beltrami operator on the unit sphere.

To calculate ${\mathcal R}_h$, one can use the transformation rule \eqref{rescale R N}
of scalar curvatures under Weyl rescalings. With the help of \eqref{R_g} it leads to
\be\label{final curvature1}
\mathcal{R}_h(x) =\left( N+M\right) \left( N+M-1\right)+
\frac{M(M-1)\,(\k^2-1)}{\r^2(x)}~.
\ee

\begin{figure}
\centering{\includegraphics[height=7cm]{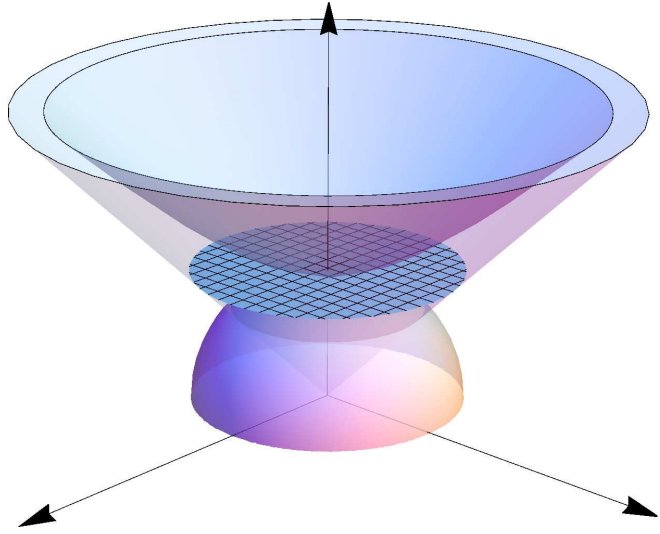}}
\caption{The unit disk here is tangent to the Lobachevski plane at the pole.
The coordinates on the disk used in the parametrization \eqref{coordinates} correspond
to the stereographic projection
onto the disk made from the
origin of the 3d space. The unit half-sphere is also tangent to the disk at the pole.
The same coordinates parameterize the unit half-sphere by the vertical projection.}
\label{plot}
\end{figure}

Now we consider the dynamical integrals related to the SO$(2,N) \times \mbox{SO}(M+1)$
isometry transformations
\be\label{integrals}
J_{AB}=\mathcal{V}_{AB}^{\,a}\,p_a +\mathcal{V}_{AB}^{\,0}\,p_0 \,,\qquad L_{IJ} =
\mathcal{V}_{IJ}^{\,\a}\,\pi_\a ~.
\ee
Here $p_a$ and $\pi_\a$ are the canonically conjugated variables to $x^a$  and $\phi^\a$, respectively,
$p_0$ is the negative square root of
\be\label{Energy square}
E^2=\frac{R^2}{\r^2}\left[g^{ab}(x)\,p_a p_b+g^{\a\b}(\phi)\,\pi_\a \pi_\b+{\mathcal M}^2\right]
\ee
and the coefficients of the momentum variables are the components of the Killing vector fields
(see \eqref{VIJ(Y)} in Appendix)
\be\label{V components A}
\mathcal{V}_{AB}^{\,0}=g^{00}(X_B\,\p_\th X_A-X_A\,\p_\th X_B)~,
\qquad \mathcal{V}_{AB}^{\,a}=g^{ab}(X_B\,\p_b X_A-X_A\,\p_b X_B)~,
\ee
\be\label{V components S}
\mathcal{V}_{IJ}^{\,\a}=g^{\a\b}(Y_J\,\p_\b Y_I-Y_I\,\p_\b Y_J)~.
\ee

Since we use dimensionless coordinates, the momentum variables $p_a,$ $\pi_\a$ and the energy are also
dimensionless.\footnote{In fact, $E^2$  corresponds
to the energy square in units of $1/ R^2$.}
The minimal energy corresponds to the vanishing momenta and the maximal value of $\r$ (i.e. $\r=1)$,
that yields $E^{\,2}_{min}={\mathcal M}^2\,R^2$.

The vector field components for the boost generators in \eqref{V components A} depend on the time
coordinate $\th$ and to use them in \eqref{integrals},
one has to make the replacement $\th \mapsto E\tau$ corresponding to the gauge fixing \eqref{gauge}.
This calculation for the boost generators at $\tau=0$ yields
\be\label{J0n=}
J_{a0'} = E x^a~, \qquad J_{a0} = (p_b\, x^b) x^a-p_a~.
\ee

From the canonical Poisson brackets $\{p_a,\,x^b\}=\d_{ab}$ follows that the energy square
\eqref{Energy square} and the boosts \eqref{J0n=} satisfy the Poisson bracket relations
\be\label{pb1}
\{E^2,\,J_{a0'}\}=-2E J_{a0}~ , \qquad \{E^2,\,J_{a0}\}=2E J_{a0'}~,
\ee
\be\label{pb2}
\{J_{a0'},\,J_{b0'}\}=-J_{ab}=\{J_{a0},\,J_{b0}\}~, \qquad \{J_{a0'},\,J_{b0}\}=E \d_{ab}~,
\ee
where $J_{ab}$ have the standard form of the rotation generators
\be\label{J_ab}
J_{ab}=p_a\,x^b-p_b\,x^a
\ee
and they correspond to the dynamical integrals \eqref{integrals} for the SO$(N)$ rotations in AdS$_{N+1}$.
Equations \eqref{pb1} and \eqref{pb2} are
then equivalent to a part of the commutation relations of the symmetry group Lie algebra.
The rest part of the algebra is trivially fulfilled, since the rotation generators both
in AdS$_{N+1}$ and S$^M$ commute with the energy square \eqref{Energy square}.
Thus, the Poisson bracket algebra of the dynamical integrals \eqref{integrals}
is not deformed by the Hamiltonian reduction,
which is a consequence of their gauge invariance  for the initial system \eqref{action g}.

Concluding this section we comment on the Casimir numbers of the isometry groups.
From \eqref{V components A} and \eqref{V components S} follows (see \eqref{VV} in Appendix) that
\be\label{Casimir numbers}
\frac{1}{2} J_{AB}J^{AB} =-R^2\left(g^{00} E^2 +g^{ab}p_a p_b\right) , \qquad
\frac{1}{2} L_{IJ} L_{IJ} =R_S^2 \,\,g^{\a\b} \pi_{\a}\pi_{\b}
\ee
and due to the mass-shell condition \eqref{massshell} we find
\be\label{Casimirs}
\frac{1}{2}\,J_{AB} J^{AB} -\k^2\,\frac{1}{2}\,L_{IJ}\,L_{IJ} ={\mathcal M}^2\,R^2~.
\ee

Notice that this constant coincides with $E^{\,2}_{min}$.

\subsection*{Quantization}

In this section we first investigate the eigenvalue problem for the energy square operator.
Then we construct the isometry group generators and check the algebra of their commutators. Finally,
we derive the quantum version of the relation between the Casimir numbers \eqref{Casimirs}.

The energy square operator is defined by \eqref{E^2 g op} and in our case its eigenfunctions
can be written in the following factorized form
\be\label{eigenfunctions}
\Psi=\psi(x)\,Y_{L}^{l_1,\cdots}(\phi)~,
\ee
where $ Y_{L}^{l_1,\cdots}(\phi)$ are the spherical harmonics on S$^M$. These states are the
eigenfunctions of the Laplace-Beltrami operator $\Delta_S$ with eigenvalues $-L(L+M-1)$ for integer $L$.

The operator $E^2$ on the AdS part then is given by
\be\label{E^2 AdS}
E^2=-h^{ab}\,\p_{ab}^2+(N+M) x^a\,\p_a+\frac{C}{\r^2}+C_1~,
\ee
with constant parameters
\be \label{C_1}
C_1=\frac{(N+M-1)^2}{4}
\ee
and $C=\mathcal{M}^2R^2+C_2+C_L$, where
\be\label{C_2}
C_2=(\k^2-1)\,\frac{(N+M-1)M(M-1)}{4(M+N)}~
\ee
corresponds to the scalar curvature term\footnote{This term vanishes at equal
radii $R=R_S$ ($\k=1$).} obtained by \eqref{R_g} and
\be\label{C_L}
C_L=\k^2\,L(L+M-1)
\ee
is the angular momentum contribution from rotations on S$^M$.
The operator \eqref{E^2 AdS} has the same structure as in the AdS space,
but with deformed parameters,
which depend on the spherical part as well. Note that the pure AdS case
corresponds to $M=0=L.$

Introducing the radial variable on the unit disk $r^2:=x^a x^a=1-\r^2$, we find $x^a\,\p_a =r\,\p_r$
and the second order derivative operator in \eqref{E^2 AdS} takes the form
\be\label{hdd}
h^{ab}\,\p_{ab}^2=\p_{aa}^2-r^2\p^2_{rr}~,
\ee
which is obviously invariant under the $SO(N)$ rotations in AdS$_{N+1}$.

Taking into account this symmetry in the operator \eqref{E^2 AdS}, one can further factorize
the eigenfunctions \eqref{eigenfunctions}
\be\label{eigenfunctions1}
\psi(x)=F(r)\,Y_l^{l_1,\cdots}(\varphi)~.
\ee
Here $Y_l^{l_1,\cdots}(\varphi)$ are again the spherical harmonics, but now on the unit sphere S$^{N-1}$
($N>1$),\footnote{The case $N=1$ will be treated separately.}
which can be treated as the boundary of the unit disk.
Using the parametrization of the radial variable
$r=\cos\s$, with $\s\in (0,\pi/2]$, and the rescaling of the radial wave function
\be\label{rescaling}
F(r)=(\sin\s)^{-\frac{M}{2}}\,(\cos\s)^{-\frac{N-1}{2}}\,f(\s)~,
\ee
we find that $f(\s)$ satisfies the Schr\"odinger equation with the P\"oschl-Teller potential
\cite{Poeschl Teller}
\be\label{radial}
\left(-\p^2_{\s\s}+\frac{A}{\sin^2\s}+\frac{B}{\cos^2\s}\right)f(\s)=E^2\,f(\s)~,
\ee
where
\be\label{A,B}
A=C+\frac{1}{4}\,M(M-2)~, \qquad B=l(l+N-2)+\frac{1}{4}\,(N-1)(N-3)~.
\ee

The integration measure for the scalar product of wave functions $f(\s)$ is just
$\,d\s$ and the Schr\"odinger operator in
\eqref{radial}, therefore, is Hermitian.

Notice that $B\geq 3/4$, except for the two cases
\be\label{case 1-2}
l=0,~~ N=3, \quad B=0~; \qquad l=0,~~ N=2, \quad B=-{1}/{4}~.
\ee

To have the Schr\"odinger operator in \eqref{radial} bounded from below (see \cite{RS}),
we assume $A\geq -{1}/{4}$.
This condition bounds the parameter $\mathcal{M}^2R^2$.

We are looking for normalizable solutions of  \eqref{radial}. To analyze them, it is convenient to
introduce the parameters
\be\label{mu,nu}
\mu:=\frac{1}{2}+\sqrt{A+\frac{1}{4}}~, \qquad \nu:=l+\frac{N-1}{2}~,
\ee
which are the large roots of the equations $\mu(\m-1)=A$ and $\n(\n-1)=B$, respectively.

The behavior of a solution of \eqref{radial} at the boundaries is given by
\be\label{boundary}
f(\s)\simeq c_1\,\s^\m+c_2\,\s^{1-\m}~,~~\s\rightarrow 0; \qquad
f(\s)\simeq d_1\,(\pi/2-\s)^\n+d_2\,(\pi/2-\s)^{1-\n}~,~~\s\rightarrow \pi/2~.
\ee
The normalizability of $f(\s)$ then requires that $c_2=0$ if $\m\geq 3/2$, and $d_2=0$ if $\n\geq 3/2$.

The case $B\geq 3/4$ is equivalent to $\n\geq 3/2$ and the corresponding solution of \eqref{radial} for
$d_1=1$ and $d_2=0$ is given by
\be\label{Solution}
f(\s)=(\sin\s)^\m\,(\cos\s)^\n\,\,_2F_1\left(a,b,c;\cos^2\s\right)~,
\ee
with the following parameters of the hypergeometric function
\be\label{parameters}
a=\frac{1}{2}\,(\mu+\nu-E)~, \qquad b=\frac{1}{2}\,(\mu+\nu+E)~,\qquad c=\nu+\frac{1}{2}~.
\ee

Since the wave function \eqref{rescaling} has to be regular at $\s=\pi/2$,
the solution \eqref{Solution} describes the two exceptional cases \eqref{case 1-2} as well.
To analyze the behavior of this solution at the boundary $\s=0$, let us consider the following
two solutions
of \eqref{radial}
\be\label{Solution 1}
f_1(\s)=(\sin\s)^\m\,(\cos\s)^\n\,\,_2F_1\left(a,b,a+b-c+1;\sin^2\s\right)~,
\ee
\be\label{Solution 2}
f_2(\s)=(\sin\s)^{1-\m}\,(\cos\s)^\n\,\,_2F_1\left(c-a,c-b,c-a-b+1;\sin^2\s\right)~,
\ee
where $f_1(\s)$ corresponds to $c_1=1,$  $\,c_2=0$ in \eqref{boundary} and
$f_2(\s)$  to $c_1=0,$  $\,c_2=1$.

The properties of hypergeometric functions then provide
\be\label{Solution=}
f(\s)=\frac{\Gamma(c-a-b)\Gamma(c)}{\Gamma(c-a)\Gamma(c-b)}\,\,f_1(\s)+
\frac{\Gamma(a+b-c)\Gamma(c)}{\Gamma(a)\Gamma(b)}f_2(\s)~.
\ee

If $\mu\geq 3/2$, the normalizability condition $c_2=0$ requires $a=-n$ with integer $n$,
and this condition leads to the energy spectrum
\be\label{energy spectrum}
E_{n,l,L}=\frac{N}{2}+\sqrt{A+\frac{1}{4}}+l+2n~.
\ee
The dependence on $L$ (and on the mass parameter and the radii)
is contained in the parameter $A$
(see \eqref{A,B} and \eqref{C_L}).

Let us now take $ \frac{1}{2}\leq\mu <\frac{3}{2}$,
which corresponds to $-\frac{1}{4}\leq A< \frac{3}{4}$.
In this case the solution \eqref{Solution} is normalizable for any (even complex) values of $E^2$,
which means that the Schr\"odinger operator in \eqref{radial} is not essentially self-adjoint.
However the analysis
of the deficiency indices \cite{RS} shows that this operator has self-adjoint
extensions. There are two different acceptable self-adjoint extensions with wave
function $f_1(\s)$ and $f_2(\s)$, which are specified
by the following boundary behavior at $\s\rightarrow 0$
\be\label{baundary 1}
f_1(\s)\simeq \s^\m~, \qquad   f_2(\s)\simeq \s^{1-\m}~.
\ee
The energy spectrum in the first case is obviously given again by \eqref{energy spectrum}
and in the second case one finds $c-b=-n$, which is equivalent to
\be\label{energy spectrum -}
E_{n,l,L}^{(-)}=\frac{N}{2}-\sqrt{A+\frac{1}{4}}+l+2n~.
\ee

For $N=1$ we use the parametrization
$x^1=-\cos\s$, with $\s\in (0,\pi),$ and the same rescaling as in \eqref{rescaling}.
This leads again to the Schr\"odinger equation \eqref{radial} with vanishing $B$
\be\label{N=1}
\left(-\p^2_{\s\s}+\frac{A}{\sin^2\s}\right)f(\s)=E^2\,f(\s)~.
\ee

If $A\geq 3/4$, the Schr\"odinger  operator in \eqref{N=1} is essentially self-adjoint
with spectrum
\be\label{N=1 spectrum}
E_n^2=(\mu+n)^2~,
\ee
where $n$ is non-negative integer and $\m$ is given by \eqref{mu,nu}.

If  $-1/4 \leq A< 3/4$ the Schr\"odinger operator in \eqref{N=1}
is not self-adjoint and one can consider two different self-adjoint
extensions similarly to the higher dimensional cases.

An interesting case here is $A=0$ (i.e. $\mu=1$), which corresponds to a free particle in a box.
The operator $-\p^2_{\s\s}$ is then characterized by two different self-adjoint extensions.
The corresponding eigenstates are give by the trigonometric functions
\be\label{sin-cos}
f_{1,\,n}(\s)=\sin(n+1)\s~, \qquad f_{2,\,n}(\s)=\cos n\s~,\quad (n\geq 0),
\ee
where $f_{1,\,n}$ and $f_{2,\,n}$  satisfy Dirichlet and Neumann boundary conditions,
respectively.

Now we discuss the isometry group generators.
The quantization of the rotation generators both on S$^M$ and AdS$_{N+1}$ is trivial.
Up to the factor $(-i)$, these operators
are given by the corresponding vector fields obtained from \eqref{V components A}-\eqref{V components S}.
Therefore, it suffices to discuss the construction of
the boost generators only.

We introduce the boost operators, which correspond to the functions \eqref{J0n=}, similarly
to the AdS case \cite{Dorn:2010wt}
\be\label{boosts 0',0}
J_{a0'}=\sqrt{E}~x^a\,\sqrt{E}~,\qquad
J_{a0}=i \sqrt{E}~\left(V_a -\frac{N+M-1}{2}\,x^a\right)~\frac{1}{\sqrt{E}}~,
\ee
with
\be\label{V_a}
V_a:=h^{ab}\p_b=\p_a-x^a x^b\p_b~.
\ee

The calculation of the commutation relations of the operators $x^a$ and $V_a$ with the energy square
operator \eqref{E^2 AdS} is straightforward (see \eqref{[E^2, x]} in Appendix) and we obtain the
quantum version of the Poisson bracket relations \eqref{pb1}
\be\label{[E^2,Jn0',0]=}
[E^2,\,J_{a0'}]=2i J_{a0}\,E+J_{a0'}~, \qquad [E^2,\,J_{a0}]=-2i J_{a0'}\,E+J_{a0}~.
\ee

Note that the value \eqref{C_1} for the constant $C_1$  is important to verify the second relation
in \eqref{[E^2,Jn0',0]=}. This confirms the value of the coefficient of the scalar curvature
term in \eqref{E^2 g op}.

From the commutation relations \eqref{[E^2,Jn0',0]=} follow the commutators
\be\label{[E,Z]}
[E, Z_a]=-Z_a~,\qquad [E, Z^*_a]=Z^*_a~,
\ee
where  $Z_a=J_{a0'}-iJ_{a0}$ and $Z^*_a=J_{a0'}+iJ_{a0}$ are the lowering and raising operators
in the ${\mathrm{so}}(2,N)$ algebra. Then it remains to calculate the commutators between
the boosts only.

In particular, by \eqref{boosts 0',0} one gets
\be\label{[J_{m0'},J_{n0'}]}
[J_{a0'},J_{b0'}]=\sqrt{E}\big(x^a\,E\,x^b\,E-x^b\,E\,x^a\,E\big)\frac{1}{\sqrt{E}}~.
\ee
To simplify the right hand side here, we apply the same trick as in \cite{Dorn:2010wt}.
Using the commutation relation $[E,J_{a0'}]=iJ_{a0}$, which follows from \eqref{[E,Z]},
one finds the operator equality
\be\label{[E,x]}
E\,x^a\,E-x^a\,E^2=\frac{N+M-1}{2}\,x^a-V_a~.
\ee
The operator in the
parentheses of \eqref {[J_{m0'},J_{n0'}]}, therefore, corresponds to the generator of
SO$(N)$ rotations \eqref{J_ab} given by $iJ_{ab}=x^b\p_a-x^a\p_b$.
Since $[E,J_{ab}]=0$, one can neglect the $\sqrt{E}$ terms
in \eqref {[J_{m0'},J_{n0'}]} and obtain
\be\label{[J,J]}
[J_{a0'}\,,\,J_{b0'}]=i J_{ab}.
\ee

The other commutators between the boost operators are computed in a similar way and they realize the algebra
\eqref{pb2} on the quantum level.

The lowering and raising operators $Z_a$ and  $Z^*_a$ provide an alternative way for calculation of the
energy spectrum.
One can start with the ground state $\psi_{0}$ of the operator \eqref{E^2 AdS}. This state is
a $\,\mbox{SO}(N)\,$ scalar and it is annihilated by the lowering operators
\be\label{Z_a}
Z_a=\sqrt{E}~\left(V_a +x^a\Big(E-\frac{N+M-1}{2}\Big)\right)~\frac{1}{\sqrt{E}}~.
\ee
The ground state wave function $\psi_{0}(\r)$ then satisfies the equations
\be\label{gs eq. N}
V_a\,\psi_{0}(\r)= x^a\left(\frac{N+M-1}{2}-E_0\right)\,\psi_{0}(\r) ~,\quad a=1,...N,
\ee
where $ E_0$ denotes the energy of the ground state. Using \eqref{V(rho)}, we find the solution
\be\label{vacuum N}
\psi_{0}(\r)=c_0\, \r^{E_0 -\frac{N+M-1}{2}}~,
\ee
up to a normalization constant $c_0$.
The normalizability condition with integration measure \eqref{det h}
restricts the minimal energy  by the unitarity bound in the AdS  space \cite{Breitenlohner:1982bm}
\be\label{unitrity bound}
 E_0 >\frac{N}{2}-1~.
\ee
The wave function \eqref{vacuum N} should also be an eigenfunction of the energy square
operator \eqref{E^2 AdS} with eigenvalue $ E_0^2$.
This condition relates $E_0$ to other parameters of the theory in the form
\be\label{mass relation}
\left(E_0 -\frac{N}{2}\right)^2=A+\frac{1}{4}~,
\ee
where $A$ is given by \eqref{A,B}. If $1/4\leq A<3/4$ one
gets two solutions of \eqref{mass relation} as above.
However, since $A$ is unbounded for increasing $L$,
one has to take only the large root $E_0^+$ of \eqref{mass relation}.
The action of the raising operators on the ground state shifts
the energy levels by one and we obtain the same spectrum as above and
the following representation of the isometry group
\be\label{representation}
\sum_{L\geq 0} D_{E_0^+(L)}(\mbox{O}(2,N))\otimes D_L(\mbox{O}(M+1))~,
\ee
where $D_{E_0^+(L)}(\mbox{O}(2,N))$ and $D_L(\mbox{O}(M+1))$ are the corresponding
standard representations.

Taking the case of equal radii ($\k=1$) and $M=N+1$, we obtain
\be\label{minimal energy}
E_0^+(L)=\frac{N}{2}+\sqrt{\mathcal{M}^2R^2+L(L+N)+\frac{N^2}{4}}~.
\ee

Finally, we consider the quantum version of the relation \eqref{Casimirs} between the Casimir numbers.
In the calculation of the Casimir number operator of the SO$(2,N)$ group one can use
the identity \eqref{[E,x]} for the term $-J_{a0'}\,J_{a0'}$ and the
factors $\sqrt{E}$  in the term $-(J_{a0'}\,J_{a0'}+J_{a0}\,J_{a0})$
can be removed by the same trick as in \eqref{[J_{m0'},J_{n0'}]}.
The computation of the remaining part is  straightforward,
and taking into account that  $\frac{1}{2}\,L_{IJ}\,L_{IJ}=-\Delta_S$,
we find
\be\label{q-Casimirs}
\frac{1}{2}\,J_{AB}\,J^{AB}-\k^2\,\frac{1}{2}\,L_{IJ}\,L_{IJ} ={\mathcal M}^2\,R^2+
C_2-\frac{N^2-(M-1)^2}{4}~.
\ee
Notice that the quantum correction here vanishes for $\k=1$ (i.e. $R=R_s$) and $N+1=M$.

\subsection*{Discussion}

A consistent quantization of a scalar particle dynamics in AdS$_{N+1}$ should provide an unitary irreducible
representation $D_{E_0}(\mbox{O}(2,N))$  of the isometry group O$(2,N)$.
This representation is characterized by the minimal value of energy $E_0$, with $\,E_0>\frac{N}{2}-1$.
The Casimir number $J^2=\frac{1}{2}\,J_{AB}\,J^{AB}$ is related to the minimal energy by $J^2=E_0(E_0-N)$ and
there are two different representations for a given $J^2$, if
\be\label{Casimir bound}
-\frac{N^2}{4}<J^2<-\frac{N^2}{4}+1~.
\ee

These are well known results and the quantization schemes mentioned in the introduction
provide various realizations of Hilbert spaces for the dynamics in AdS backgrounds.

The static gauge quantization leads to position-dependent wave functions, which
seems to be most natural. However, the relativistic principles require a non-local character for
the energy and boosts operators, that creates complicated ordering ambiguities.
Due to this ordering problem the static gauge quantization was usually avoided in the literature.
The construction of the energy square operator, proposed in \cite{Dorn:2010wt},  simplifies the ordering
ambiguity problem and allows to solve it in AdS$_{N+1}$.

The main result of the present paper is the generalization of the static gauge quantization to the AdS$\times$S spaces.
This result is summarized in equations \eqref{energy spectrum},
\eqref{mass relation}-\eqref{q-Casimirs}.

It is remarkable that the Hamiltonian of the system in the static gauge coincides with the energy square,
which is characterized by the classical parameters $\mathcal{M}$, $R$ and $R_S$. Upon quantization
the quantum minimal energy and the quantum relation between the
Casimir numbers are expressed through these classical parameters.
These expressions exhibits the quantum deformations of the corresponding classical relations $E_{min}=\mathcal{M}R=\sqrt{J^2-\k^2 L^2}$.

Notice that the connection to the classical mass parameter sometimes is lost in various quantization schemes.

An interesting new observation is the existence of two different self-adjoint extensions of the Hamilton operator
for those values of the parameter $\mathcal{M}R$ which correspond to \eqref{Casimir bound}.
This gives a new interpretation to two non-equivalent representations of the O$(2,N)$ isometry
group for a given Casimir number $J^2$.

In pure AdS space, which corresponds to $M=0=L$, we reproduce the results of \cite{Dorn:2010wt}
for arbitrary AdS$_{N+1}$.
Note that the energy spectrum of the AdS$_5$ particle obtained in \cite{Claus:1999jj} with an additional
requirement of the CPT symmetry, corresponds the Breitenlohner-Freedman bound \cite{Breitenlohner:1982bm}
with $E_0=2$ and $J^2=-4$.

The most interesting case is AdS$_5\times$S$^5$ with equal radii, where the minimal energy is given by
\eqref{minimal energy}.
At large $\mathcal{M}R$ one gets the following expansion
\be\label{expansion of minimal energy}
E_0=\mathcal{M}R+2+\frac{L(L+4)+4}{2\mathcal{M}R} +\mathcal{O}((\mathcal{M}R)^{-2})~.
\ee
This result was obtained in \cite{Passerini:2010xc} on the basis of Dirac bracket quantization and it has been used
for the analysis of the string energy spectrum in perturbation theory at large coupling.

Finally we would like to comment on the static gauge quantization of string dynamics.
The ordering problem for the Poincare group generators in Minkowski space has been solved in \cite{Jorjadze:2012iy}
similarly to the particle case and the critical dimension
$D=26$ has been reproduced by the closure of the Poincare algebra.
The application of the same scheme to the AdS background provides a convenient framework for calculation
of the string spectrum in perturbation theory at strong coupling and the work in this direction is in progress.

\vspace{3mm}

\noindent
{\bf Acknowledgments}

\vspace{3mm}

\noindent
We thank Harald Dorn for his interest in our work and for useful comments.
G.J. thanks Martin Reuter for helpful discussions on static gauge quantization.
This work was done during the visit of G.J. and Z.K. at Humboldt University of Berlin.
They thank the department of Physics for warm hospitality.

\noindent
This work has been supported in part by the grant I/84600 from VolkswagenStiftung.

\setcounter{equation}{0}

\def\theequation{A.\arabic{equation}}

\subsection*{Appendix}

In this Appendix we collect some useful formulae used in the main text.

\vspace{3mm}

\noindent
{\bf Weyl transformation of a scalar curvature}

\vspace{3mm}

\noindent
If a metric tensor $h_{mn}(x)$ is a Weyl transformed $g_{mn}(x)$
$$h_{mn}(x) = \frac{1}{\L(x)}\,g_{mn}(x)~,$$
then its scalar curvature is given by
\be\label{rescale R N}
{\mathcal{R}}_h=
\L\,\left(\mathcal{R}_{g}+(\mathcal{N}-1)\Delta_{g}(\log\L)-
\frac{(\mathcal{N}-1)(\mathcal{N}-2)}{4}\,\,g^{mn}\,\p_m(\log\L)\,\p_n(\log\L)\,\right)~.
\ee

\noindent
{\bf Killing vector fields on AdS$_{N+1}$ and S$^M$}

\vspace{3mm}

\noindent
The Killing vector fields on AdS$_{N+1}$ and S$^M$ are treated similarly and it is sufficient to consider
the spherical part only.

Let us introduce vector fields $\mathcal{V}_I$ on S$^M$ with components $\mathcal{V}_I^\a=g^{\a\b}\p_\b Y_I$.
These fields are obtained from the equations $\mathcal{V}_I\,\rfloor g= \mathrm{dY}_I~$ and they satisfy the relations
\be\label{V(Y)}
\mathcal{V}_{I}(Y_K)=\d_{IK}-\frac{Y_I\,Y_K}{R^2}~.
\ee
The Killing vector fields $\mathcal{V}_{IJ}$ with components \eqref{V components S}
then can be written as $\mathcal{V}_{IJ}=Y_J\,\mathcal{V}_I-Y_I\,\mathcal{V}_J,$ and due to \eqref{V(Y)} they provide
the infinitesimal form of the isometry transformations
\be\label{VIJ(Y)}
\mathcal{V}_{IJ}(Y_K)=\d_{IK}\,Y_J-\d_{JK}\,Y_I~.
\ee
The metric on S$^M$ can be written as $\mathrm{dY}_I\,\mathrm{dY}_I=\frac{1}{2}\,\th_{IJ}\,\th_{IJ}$, with
$\th_{IJ}=Y_I\,\mathrm{dY}_I-Y_I\,\mathrm{dY}_J$, and in local coordinates one finds
\be\label{VV}
\frac{1}{2}\,\mathcal{V}_{IJ}^\a\,\mathcal{V}_{IJ}^\b=g^{\a\b}~.
\ee

\vspace{3mm}

\noindent
{\bf Algebra of the boost operators}

\vspace{3mm}

\noindent
The energy square operator \eqref{E^2 AdS} is linear in $H=-h^{ab}\p^2_{ab}$ and $D=x^a\p_a$, whereas the
boost generators \eqref{boosts 0',0} are linear in $V_a$ and in the multiplication operator $x^a$.
The commutators of these operators form the following algebra
\bea\label{[D,x]}
&& [D,x^a]=x^a,\qquad \qquad~~~~~~~~ [H,x^a]=-2V_a \\ \nonumber
&& [D,V_a]=-2x^a D-V_a~,\qquad  [H,V_a]=2V_a-2x^a H~,
\eea
and, in addition, one has
\be\label{V(rho)}
[V_a,\r]=V_a(\rho)=-x^a\rho~.
\ee
From these commutators follow
\be\label{[E^2, x]}
[E^2,x^a]=-2V_a+(M+N)x^a~,\qquad [E^2,V_a]=-2x^a(E^2-C_1)-(M+N-2)V_a~,
\ee
which turn the equations in \eqref{[E^2,Jn0',0]=} to identities.

\end{document}